\documentclass[preprint]{aastex}
\usepackage{psfig}

\slugcomment{To appear in\it The Astrophysical Journal, Letters}

\shorttitle{X-Ray-Luminous AGN in Mrk 231}
\shortauthors{Maloney \& Reynolds}

\begin{document}

%USER-DEFINED DEFINITIONS

\def\ie{{\it i.e.,\ }}
\def\eg{{\it e.g.,\ }}
\def\qv{{\it q.v.,\ }}
\def\cf{{\it cf.\ }}
\def\etal{{\it et al.}}
\def\gtrapprox{\;\lower 0.5ex\hbox{$\buildrel >
    \over \sim\ $}}             %greater than about
\def\lessapprox{\;\lower 0.5ex\hbox{$\buildrel < \over \sim\ $}}
\def\msol{\ifmmode {\>M_\odot}\else {$M_\odot$}\fi}
\def\Lsol{\ifmmode {\>L_\odot}\else {$L_\odot$}\fi}
\def\Zsol{\ifmmode {\>Z_\odot}\else {$Z_\odot$}\fi}
\def\pyr{\ifmmode {\>{\rm\ yr}^{-1}}\else {yr$^{-1}$}\fi}
\def\psec{\ifmmode {\>{\rm\ s}^{-1}}\else {s$^{-1}$}\fi}
\def\kms{\ifmmode {\>{\rm km\ s}^{-1}}\else {km s$^{-1}$}\fi}
\def\pcmsq{\ifmmode {\>{\rm cm}^{-2}}\else {cm$^{-2}$}\fi}
\def\pcubcm{\ifmmode {\>{\rm cm}^{-3}}\else {cm$^{-3}$}\fi}
\def\hubunits{\ifmmode {\>{\rm km\ s^{-1}\ Mpc^{-1}}}\else {km s$^{-1}$ 
 Mpc$^{-1}$}\fi} 
\def\Lx{\ifmmode {L_x}\else {$L_x$}\fi}
\def\Lbol{\ifmmode {L_{bol}}\else {$L_{bol}$}\fi}
\def\Lir{\ifmmode {L_{IR}}\else {$L_{IR}$}\fi}
\def\ergps{\ifmmode{{\rm erg\, s^{-1}}}\else {erg s$^{-1}$}\fi}
\def\eV{{\rm\thinspace eV}}
\def\keV{{\rm\thinspace keV}}
\def\be{\begin{equation}}
\def\ee{\end{equation}}
\def\bea{\begin{eqnarray}}
\def\eea{\end{eqnarray}}

\title{{\sl ASCA} Observation of an X-Ray-Luminous Active Nucleus in
Markarian 231}

\author{Philip R. Maloney\altaffilmark{1}}
\affil{Center for Astrophysics \& Space Astronomy, 
University of Colorado,
Boulder, Colorado, 80309-0389}
\altaffiltext{1}{maloney@casa.colorado.edu}
\and
\author{Christopher S. Reynolds\altaffilmark{2,3}}
\affil{JILA, 
University of Colorado,
Boulder, Colorado, 80309-0440}
\altaffiltext{2}{chris@rocinante.colorado.edu}
\altaffiltext{3}{Hubble Fellow}

\begin{abstract}
We have obtained a moderately long ($10^5$ second) {\it ASCA}
observation of the Seyfert 1 galaxy Markarian 231, the most luminous
of the local ultraluminous infrared galaxy (ULIRG) population. In the
best-fitting model we do not see the X-ray source directly; the
spectrum consists of a scattered power-law component and a reflection
component, both of which have been absorbed by a column $N_H\approx
3\times 10^{22}\pcmsq$. About $3/4$ of the observed hard X-rays arise
from the scattered component, reducing the equivalent width of the
iron K$\alpha$ line. The implied ratio of $1-10$ keV X-ray luminosity
to bolometric luminosity, $\Lx/\Lbol\sim 2\%$, is typical of Sy 1
galaxies and radio-quiet QSOs of comparable bolometric luminosities,
and indicates that the bolometric luminosity is dominated by the
AGN. Our estimate of the X-ray luminosity also moves Mrk~231 in line
with the correlations found for AGN with extremely strong Fe II
emission. A second source separated by about 2 arcminutes is also
clearly detected, and contributes about $25\%$ of the total flux.
\end{abstract}


\section{Introduction}
\label{introduction}
Sixteen years ago, the {\sl IRAS} all-sky survey revealed the
existence of a class of objects which emit nearly all of their energy
at far-infrared wavelengths (Soifer \etal\ 1984). At the
high-luminosity end of the distribution, $\Lir\ge 10^{12}\Lsol$, the
so-called ``ultraluminous infrared galaxies'' (ULIRGs), the space
density of such objects is approximately twice that of
optically-selected QSOs, the only objects with comparable bolometric
luminosities (Sanders \& Mirabel 1996). The far-infrared energy
distributions of the prototype objects, Arp~220 and NGC 6240, are very
similar to that of the classic starburst galaxy M82, peaking at
$\lambda\simeq 60\mu$m, and do not exhibit the ``warm'' {\sl IRAS}
colors typical of Seyfert galaxies (DeGrijp \etal\ 1992). A large
fraction of the ultraluminous galaxies show evidence for interaction
or merging with a companion galaxy (Sanders \etal\ 1988; Melnick \&
Mirabel 1990), and most show luminous molecular line emission,
indicating substantial quantities of molecular gas, which could
potentially fuel either a burst of star formation or an AGN (\eg
Sanders \& Mirabel 1996, and references therein).

Ever since the discovery of this class of galaxies debate has raged
over the nature of the dominant source of energy. The failure to
detect any of the ULIRGs at X-ray wavelengths with HEAO-1 (Rieke 1988)
strengthened a starburst interpretation, as did the recent failure of
the {\sl OSSE} instrument on the {\sl GRO} to detect Arp~220, Mrk~231,
or Mrk~273 in the $50-200$ keV energy range (Dermer \etal\ 1997); this
rules out very luminous {\sl Compton-thin} sources in these
galaxies. Most recently, observations of a number of ULIRGs with the
short-wavelength ({\sl SWS}) spectrometer on the {\sl Infrared Space
Observatory} failed to detect any high-excitation lines in the
mid-infrared, as might be expected from even a dust-shrouded AGN
(Sturm \etal\ 1996; Genzel \etal\ 1998), and this has been taken to be
evidence for starbursts rather than AGN. However, it is now clear that
in many ULIRGs the column densities toward the nucleus are so high
that the dust optical depth is greater than unity even in the
mid-IR. Furthermore, the discovery of Compton-thick X-ray sources in a
number of ULIRGs, notably including NGC 6240 (Iwasawa \& Comastri
1998; Vignati \etal\ 1999), substantially weakens the argument against
AGN-dominated sources based on hard X-ray non-detections. Such
sources, in which even hard X-rays are seen only through reflection,
are characterized by a relatively flat continuum above $E\sim 3$ keV
and very large equivalent widths of the iron $K\alpha$ line, exceeding
1 keV (\eg Krolik \& Kallman 1987; Krolik, Madau \& \.Zycki 1994;
Matt, Brandt, \& Fabian 1996). Such large equivalent widths can be
produced only if the X-ray continuum incident on the line-emitting
region is much larger than that which we observe. If the reflected
fraction in NGC 6240 is similar to that in NGC 1068, the bolometric
luminosity of the AGN is of order $10^{45}$ \ergps, and dominates the
far-infrared output of the galaxy.

Foremost among the local ULIRGs is Markarian 231 ($z=0.042$) , which
is the most luminous object in the local universe ($z < 0.1$) (Soifer
\etal\ 1984). Its bolometric luminosity is $L_{bol}\approx 3\times
10^{12}\Lsol$ ($H_o=75$ km s$^{-1}$ Mpc$^{-1}$), nearly all of which
emerges in the far-infrared. Although classified as a Seyfert 1,
Mrk~231 shows evidence for substantial amounts of reddening ($A_V\sim
2$: Boksenberg \etal\ 1977) and when corrected for extinction its
optical luminosity approaches quasar levels. Furthermore, Mrk~231
exhibits many features characteristic of (low ionization) Broad
Absorption Line (BAL) QSOs: high optical polarization ($20\%$ at
2800\AA, Smith \etal\ 1995), strong Fe II emission (extremely strong in
Mrk~231) (Boksenberg \etal\ 1977), red optical colors, and a large
IR/optical flux ratio. {\sl ASCA} and {\sl ROSAT} observations of
Mrk~231, analyzed by Iwasawa (1999) and Turner (1999), revealed an
extended, soft X-ray component, with a luminosity $L\sim
10^{42}\ergps$, and a hard component exhibiting considerable
absorption, with a 2-10 keV luminosity of only $L\approx 2\times
10^{42}\ergps$. If this is the intrinsic X-ray luminosity, then an AGN
is unlikely to dominate the energetics unless Mrk~231 is
X-ray-quiet. However, the signal-to-noise of the {\sl ASCA} spectrum
was too low to determine whether the observed X-ray emission is
largely reflected, in which case the true X-ray luminosity could be
much higher.

In this letter, we present an analysis of a much deeper ASCA
observation of Mrk 231 than that presented by Iwasawa. We find strong
evidence for a powerful, buried AGN in this ULIRG which we observe
only through reflection and scattering. Indeed, we show that the AGN
in Mrk231 may well be the dominant energy source.

\section{Observation and Data Reduction}
\label{observation}
Mrk~231 was observed by {\it ASCA} on 1999-Nov-10. The SIS data were
cleaned in order to remove the effects of hot and flickering pixels
and subjected to the following data-selection criteria : i) the
satellite should not be in the South Atlantic Anomaly (SAA), ii) the
object should be at least 7 degrees above the Earth's limb, iii) the
object should be at least 25 degrees above the day-time Earth limb,
and iv) the local geomagnetic cut-off rigidity (COR) should be greater
than 6\,GeV/c. We also applied a standard grade selection on SIS
events in order to further minimize particle background. The GIS data
were cleaned to remove the particle background and subjected to the
following data-selection criteria : i) the satellite should not be in
the SAA, ii) the object should be at least 7 degrees above the Earth's
limb and iii) the COR should be greater than 7\,GeV/c. After these
data selection criteria were applied, there are 96.3\,ksec of good
data per SIS detector and 99.2\,ksec of good data per GIS detector.

Full-band images were extracted from each of the four {\it ASCA}
detectors. Two point-like X-ray sources were found in the SIS image
separated by 2\,arcmin, with the dominant source (containing about
75\% of the counts) coincident with Mrk~231. The GIS has insufficient
spatial resolution to distinguish these sources. We extracted the SIS
spectrum of each source and found them to have very similar shapes.
Given that the fainter source does not dominate at any energy, and
that detailed spectroscopy on these small scales is difficult due to
the energy-dependent point spread function of ASCA, we chose to
extract and analyze the combined spectrum of these sources (which is
dominated by Mrk~231).

Source photons were extracted from a circular region centered on the
two sources with radii of 3\,arcmin and 4\,arcmin for the SIS and GIS
respectively. Background spectra were obtained from blank regions of
the same field (using the same chip in the case of the SIS). No
temporal variability was observed. In order to facilitate $\chi^2$
spectral fitting, all spectra were rebinned so as to contain at least
20 photons per spectral bin. In order to avoid poorly calibrated
regions of the spectrum, the energy ranges considered were
1.0--10\,keV for the SIS detectors, and 0.8--10\,keV for the GIS
detectors. Note that we use a lower-energy cutoff for the SIS that is
considerably higher than the `standard' 0.6\,keV cutoff in order to
avoid the effects of ``residual dark current distribution'', or RDD,
which is known to plague recent {\it ASCA} observations.

\section{Spectral Analysis}
\label{analysis}
Iwasawa (1999) finds evidence for soft X-ray emission from thermal
plasma and measures a temperature of $kT=0.98^{+0.51}_{-0.09}\keV$ and
an abundance of $Z=0.08^{+0.20}_{-0.05}\Zsol$. Since we deem our soft
($E<1\keV$) SIS data to be unusable, we cannot constrain the nature of
the thermal plasma component. Therefore, in all that follows, we
constrain the parameters of the thermal component ($kT$ and $Z$) to
the 90\% error range found by Iwasawa.

Guided by Iwasawa (1999), we initially fit our SIS \& GIS spectrum
(Figure 1) with a model consisting of this thermal component and a
power-law tail with photon index $\Gamma$ (representing the AGN
emission), all absorbed by the Galactic column of $N_{\rm
H}=1.26\times 10^{20}\pcmsq$. Although this spectral model produced an
adequate goodness of fit ($\chi^2/{\rm dof}=304/305$), the resulting
photon index was extremely flat ($\Gamma=0.69\pm 0.12$). In
particular, it is much flatter than the intrinsic spectrum of any
known AGN (which usually have $\Gamma=1.5-2.2$). Adding intrinsic
absorption to the power-law component increases the permitted photon
index by $\Delta\Gamma\sim 0.4$ (with $\Delta\chi^2=7$ for 1 new
parameter) --- insufficient to bring $\Gamma$ into the acceptable AGN
range.

Since this AGN is likely to be highly obscured, we also examined
classes of models in which the central power-law is completely
absorbed by a Compton-thick screen, and is only seen via scattering
and Compton reflection. In practice, we used the {\sc pexrav} models
within the {\sc xspec} package, interpreting finite reflection
fractions as representing mixed reflection/scattering (\ie scattering
of non-reflected as well as reflected continuum into our line of
sight). A model consisting of this reflected/scattered component and
the thermal plasma produced a good fit ($\chi^2/{\rm dof}=284/304$)
and implied a reflection-dominated spectrum (with the reflection
fraction ${\cal R}>5800$). However, the implied intrinsic power-law
was much steeper ($\Gamma=3.01^{+0.12}_{-0.28}$) than the normal AGN
range.

%Physically acceptable results are produced when one includes
%absorption of the reflected/scattered component. While this absorption
Physically acceptable results are produced when
absorption of the reflected/scattered component is included. While
this absorption 
is not statistically required, including a column of $N_{\rm H}\sim
3\times 10^{22}\pcmsq$ extends the range of allowable photon indices
into the typical AGN range. In order to better constrain the fits, we
hereafter assume that the hard X-ray photon index of the source is
typical of radio-quiet AGN, $\Gamma=1.8$. We also include a narrow
iron K$\alpha$ emission line of cold iron which is expected to
accompany the X-ray reflection. The best fitting parameters (with 90\%
error range) are then: ${\cal R}=8.7^{+7.9}_{-5.4}$, $N_{\rm
H}=2.9^{+2.5}_{-1.0}\times 10^{22}\pcmsq$, and an iron line equivalent
width of $W_{K\alpha}=290^{+190}_{-170}\eV$ (with $\chi^2/{\rm
dof}=280/303$); the resulting model spectrum is shown in Figure 2. We
have experimented with fits assuming partial covering of the source or
ionized rather than neutral absorbers; neither the quality nor the
nature of the best fits differ in any significant way from the
above. This is to be expected since the hard X-ray spectrum largely
decouples in the analysis from the soft, absorbed, spectrum.

We also investigated the potential presence of an unabsorbed power-law
component (corresponding to AGN emission that has been scattered at
sufficiently large distances to be free of any substantial
absorption). No such component was required by the data, and limits
were set on the luminosity of any such component of $<1.1\times
10^{42}\ergps$. 
 
\section{Discussion}
\label{discussion}
In his analysis of previous {\it ASCA} data, Iwasawa (1999) was unable
to distinguish between a very flat, unabsorbed power-law and a heavily
absorbed, typical ($\Gamma=1.8$) AGN spectrum. With the considerably
better S/N provided by our {\it ASCA} observation, we can show
unequivocally that no amount of neutral absorption can steepen the
intrinsic photon index into the normal AGN range. We are therefore led
to conclude that there is a Compton-thick screen blocking the X-ray
source along our line of sight, and we see the AGN only by reflection
and scattering of the X-rays into our line of sight (Figure
3). Furthermore, unless the spectrum of the hard X-ray emission is
unusually steep, the reflected/scattered emission is fairly heavily
absorbed, by a neutral column $N_{\rm H}\approx 3\times
10^{22}\pcmsq$.

We can estimate the intrinsic luminosity of the X-ray source in two
ways. When we subtract the contribution from the thermal emission, the
observed $0.5-10$ keV luminosity is $2.0\times 10^{42}\,\ergps$.
Corrected for neutral absorption, this becomes $3.2\times
10^{42}\,\ergps$, and with the reflection fraction ${\cal R}$ set equal
to zero, the scattered luminosity is $L_{\rm sc}=2.4\times
10^{42}\,\ergps$. (Note that this implies that about three-quarters of
the {\it observed} X-ray luminosity is due to scattering of X-rays into our
line of sight, rather than reflection, thus reducing the equivalent
width of the Fe K$\alpha$ line from the value of $\sim 1$ keV expected
for a reflection-dominated spectrum.) Hence the intrinsic luminosity
of the X-ray source is given by 
\be L_{\rm intr}=2.4\times 10^{43}\left({0.1\over f_{\rm sc}}\right)\
\ergps 
\ee
where the scattering fraction $f_{\rm sc}$ is expected to be on the
order of $1\%$ for electron scattering, assuming that the scattering
column density is of the same order as the inferred absorption column
density, \ie a Thompson optical depth of order $0.01$. 

Alternatively, we can estimate $L_{\rm intr}$ from the reflected
fraction and modeling of the reflection process. From the above
numbers, the observed luminosity in the reflection component is
$L_{\rm refl}=8.0\times 10^{41}\ergps$. Using {\sc xspec}, we conclude
that the product of $L_{\rm intr}$ and the reflection fraction $f_{\rm
refl}$ is
\be
L_{\rm intr}f_{\rm refl}=2.0\times 10^{43}\ \ergps
\ee
and so
\be
L_{\rm intr}=2.0\times 10^{44}\left({0.1\over f_{\rm refl}}\right)\
\ergps 
\ee
where we have normalized to $0.1$ under the assumption that only a
small fraction of the reflecting surface is visible to us; for the
Compton-thick Sy 2 NGC 1068, Iwasawa, Fabian \& Matt (1997) infer
$f_{\rm refl}\sim 0.05$. These two estimates of the luminosity agree
for a ratio $f_{\rm refl}/f_{\rm sc}\approx 10$, \ie for a reflected
fraction $f_{\rm refl}\approx 0.1$ for a scattered fraction $f_{\rm
sc}\approx 1\%$ which, as noted above, is the value expected if the
scattering column is of the same order as the inferred absorbing
column. (This ratio of $f_{\rm sc}/f_{\rm refl}$ is {\it not}
independent of the value ${\cal R}\approx 10$ derived from the
spectral fitting.) Thus we infer that $f_{\rm refl}\approx 0.1$,
$f_{\rm sc}\approx 0.01$, and $\Lx\approx 2\times
10^{44}\,\ergps$. This is $200$ times larger than the value of Iwasawa
(1999), and a factor $\sim 20$ larger than the luminosity estimated by
Turner (1999); hence our analysis implies that the AGN is much more
X-ray-luminous than previously thought.

From our estimate of the intrinsic hard X-ray luminosity \Lx, we can 
estimate the contribution of the AGN to the bolometric luminosity of
Mrk 231. The implied ratio of $\Lx/\Lbol\sim 2\%$. For comparison, the
ratio of $\Lx/\Lbol$ for the sample of Sy 1 and radio-quiet QSOs of
Elvis \etal\ (1994) for which X-ray data are available ranges from
$1.95\%$ to $14.5\%$, with the mean value decreasing somewhat with
\Lbol, from $6.3\%$ for $\Lbol=10^{45}-10^{46}$ (11 objects) to
$4.3\%$ for $\Lbol=10^{46}-2.2\times 10^{46}$ (7 objects); all the
latter are classified as radio-quiet QSOs. Hence, within the
uncertainties, the energetics of Mrk 231 are consistent with complete
domination by the AGN; in any case, the fraction of the bolometric
luminosity contributed by the AGN is almost certainly large
($\gtrapprox 0.5$). Extrapolating to high energies, the predicted
$50-200$ keV luminosity is $L_{50-200}\approx 2\times
10^{44}\,\ergps$, a factor of five below the Dermer \etal\ (1997) upper
limit from {\it OSSE}. 

We also note that, with our estimate of the intrinsic X-ray
luminosity, Mrk~231 is no longer an outlier in the significant
correlations found by Lawrence \etal\ (1997) in the properties of AGN
with extreme Fe II emission, in particular, those between the 2 keV --
1$\mu$m spectral index $\alpha_{\rm ix}$ and the Fe II emission
strength and the emission line FWHM. Since our estimate of the flux
density at 2 keV is $\sim 270$ times larger than the value used by
Lawrence \etal\ (from Rigopoulou, Lawrence, \& Rowan-Robinson 1996),
we find a value of $\alpha_{\rm ix}\approx 1.2$, smaller by $0.8$ than
the value inferred by Lawrence \etal, which moves Mrk~231 into line
with the bulk of strong Fe II emitters in their sample. Furthermore,
the expected hard X-ray luminosity from the correlation between broad
H$\alpha$ and $2-10$ keV emission (Ward \etal\ 1988), using the
H$\alpha$ flux measured by Smith \etal\ (1995, uncorrected for
extinction), is $L_x\gtrapprox 2\times 10^{44}\ergps$, in excellent
agreement with the inferred X-ray luminosity.

\section{Summary}
\label{summary}
A moderately long {\it ASCA} observation of the infrared-luminous
galaxy Mrk~231 shows a very flat hard X-ray spectrum, which strongly
suggests that the central AGN is hidden behind a Compton-thick screen,
and is observed in X-rays only by scattering and reflection of the
radiation into our line of sight. The inferred X-ray luminosity in the
$0.5-10$ keV band is $\Lx\approx 2\times 10^{44}\,\ergps$ for a
scattering fraction $f_{\rm sc}\approx 1\%$, the value expected from
the absorbing column density inferred from the X-ray spectrum. This is
much larger than previous estimates, and the
resulting ratio of $\Lx/\Lbol$ is typical of luminous Sy 1 and
radio-quiet QSOs, indicating that the AGN dominates the bolometric
luminosity of the galaxy, and confirming the suggestion of Smith
\etal\ (1995) that Mrk~231 is most properly classified as a
low-ionization BAL QSO. The increase in the inferred X-ray luminosity
over previous estimates also puts Mrk~231 in line with the
correlations between 2 keV -- 1$\mu$m spectral index and both Fe II
emission strength and emission line FWHM found by Lawrence \etal\ 
(1997) for AGN with extremely strong Fe II emission. 

\acknowledgements{PRM is supported by the NSF under grant AST-9900871 and
by the NASA Astrophysical Theory Program under grant NAG5-4061.
CSR appreciates support from Hubble Fellowship grant HF-01113.01-98A.  
This grant was awarded by the Space Telescope Institute, which is operated
by the Association of Universities for Research in Astronomy, Inc., for
NASA under contract NAS 5-26555. CSR also appreciates support from NASA
under LTSA grant NAG5-6337, and the National Science Foundation under
grants AST-9529170 and AST-9876887. We are grateful to Mike Nowak for
producing Figure 3.}

%\newpage

\newpage

\setcounter{figure}{0}
\begin{figure}
\centerline{\psfig{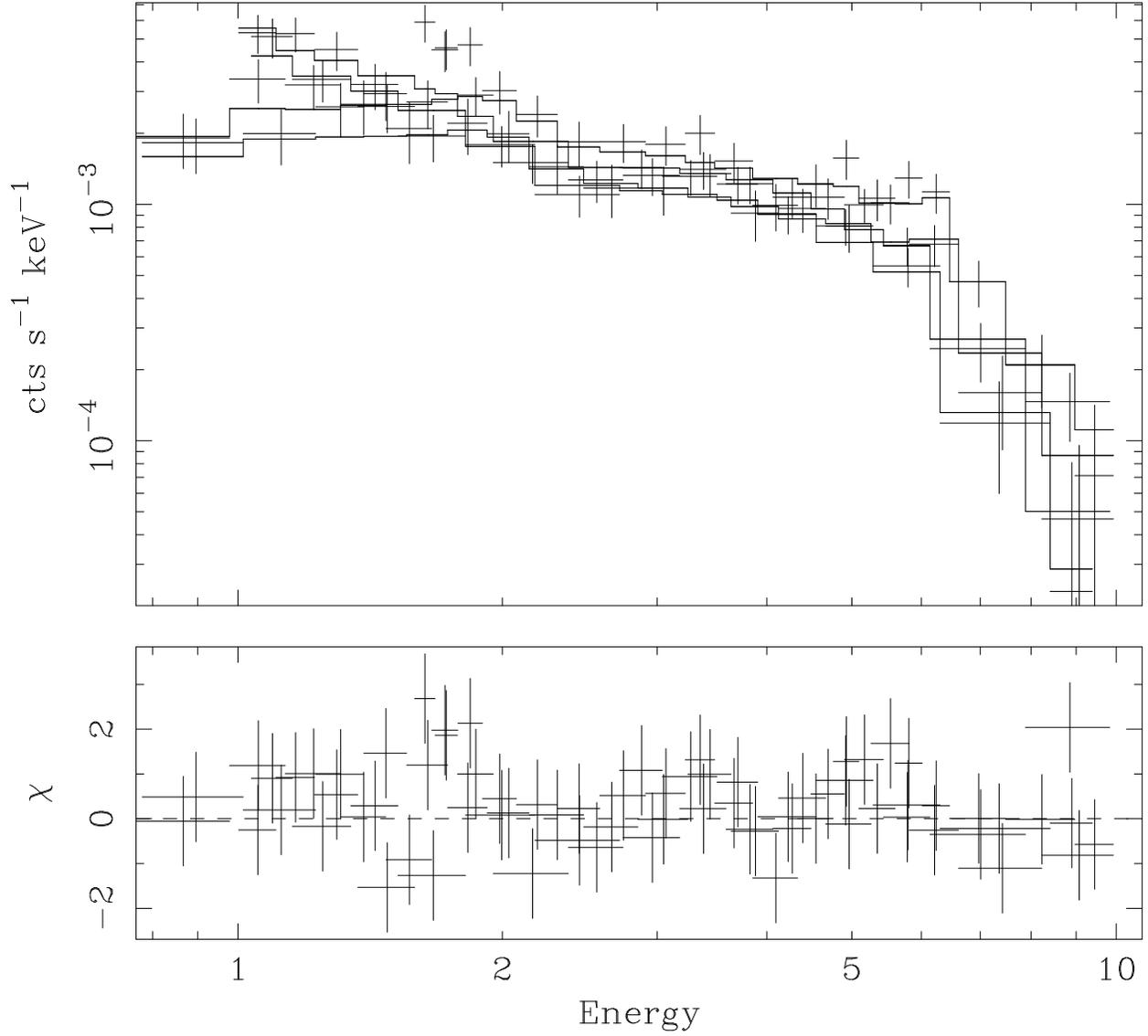}}
\caption{ASCA SIS and GIS spectrum of Mrk~231.  The top panel shows
%\figcaption{ASCA SIS and GIS spectrum of Mrk~231.  The top panel shows
the data for all four instruments together with the best fitting
folded model (i.e. the model has been folded through the response of
the telescope/detector system). The bottom panel shows the
contribution of each spectral bin to the overall $\chi^2$ of the best
fit.}
\end{figure}

\clearpage
\begin{figure}
\centerline{\psfig{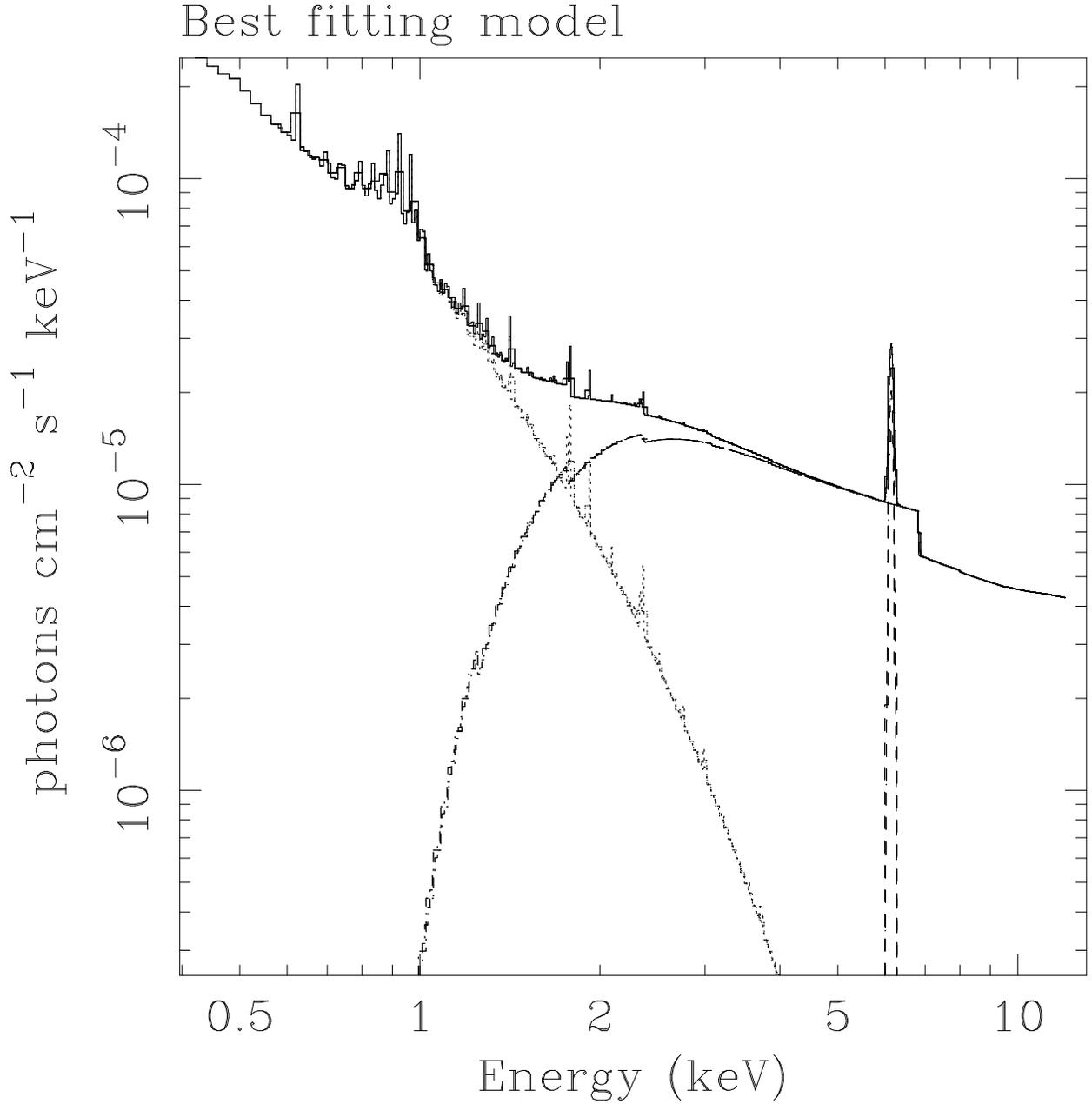}}
\caption{Best fitting spectral model for Mrk~231, plotted in terms of
%\figcaption{Best fitting spectral model for Mrk~231, plotted in terms of
photon flux. The reflected/scattered emission from the AGN component
dominates at hard X-ray energies, but is truncated at 1--2\,keV by
heavy neutral absorption. The line-rich thermal plasma spectrum
dominates at soft X-ray energies. The strong iron-L complex at $\sim
1$\,keV is clearly visible.}
\end{figure}

\clearpage
\begin{figure}
\centerline{\psfig{figure=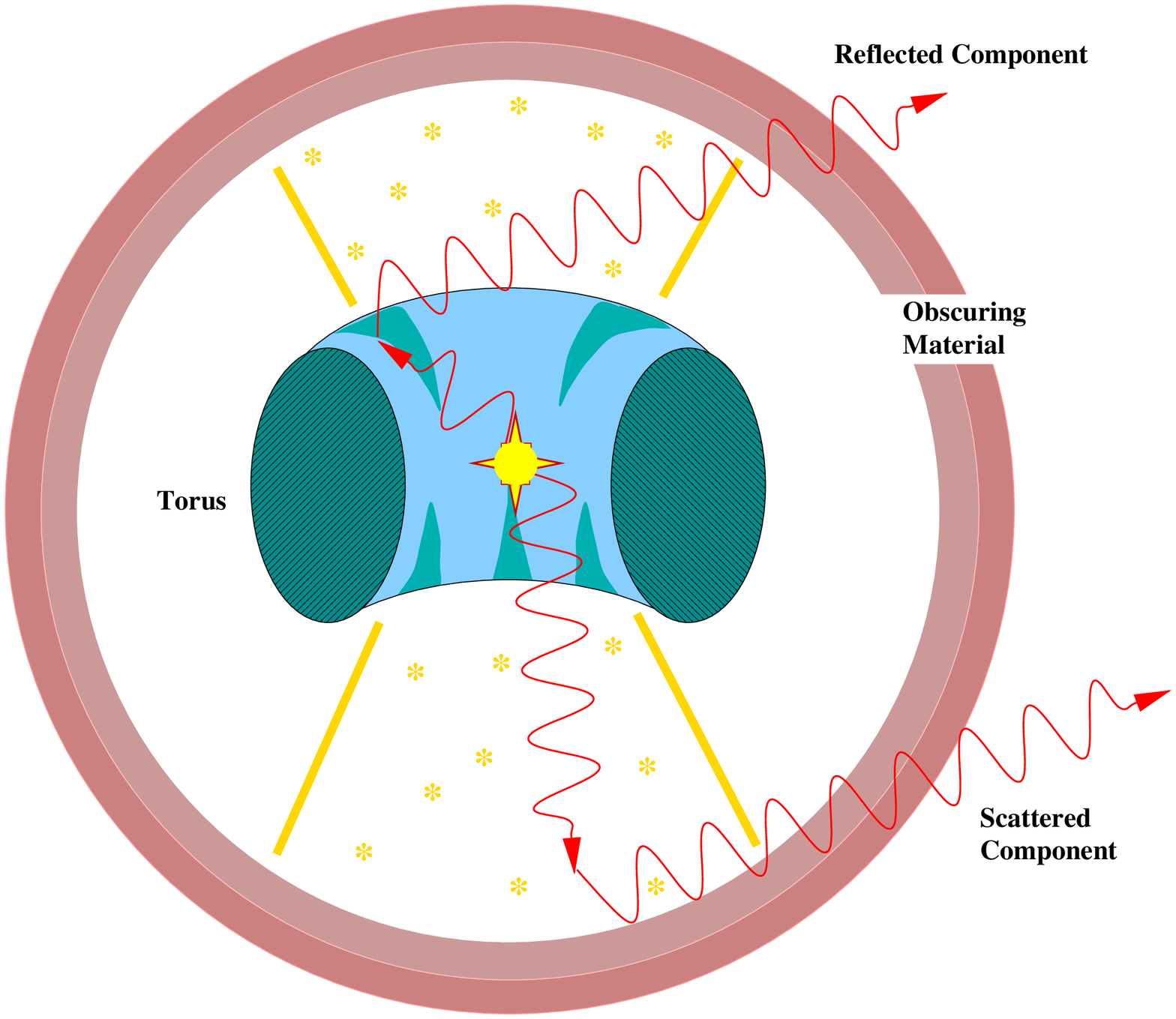,width=\textwidth,angle=0}}
\caption{Schematic illustration of our inferred geometry for the X-ray
%\figcaption{Schematic illustration of our inferred geometry for the X-ray
emission from Mrk~231, in which we observe only scattered and
reflected emission and do not have an unobstructed view of the
nucleus, and there is a distribution of absorbing gas along our line
of sight to both the reflected and scattered components of the X-ray
emission.}
\end{figure}


\begin{thebibliography}{}

\bibitem[boxy]{boks77}
Boksenberg, A., Carswell, R.F., Allen, D.A., Fosbury, R.A.E., Penston,
M.V., \& Sargent, W.L.W. 1977, MNRAS, 178, 451 

\bibitem[elvis]{elv94}
Elvis, M., Wilkes, B.J., McDowell, J.C., Green, R.F., Bechtold, J.,
Willner, S.P., Oey, M.S., Polomski, E.,\& Cutri, R. 1994, ApJS, 95, 1

\bibitem[degri]{degr92}
DeGrijp, M.H.L., Keel, W.C., Miley, G.K., Goudfrooij, P., \& Lub,
J. 1992, A\&AS, 96, 389

\bibitem[derm97]{derm97}
Dermer, C.D., Bland-Hawthorn, J., Chiang, J., \& McNaron-Brown,
K. 1997, ApJ, 484, L121 

\bibitem[genzet98]{genzet98}
Genzel, R., \etal\ 1998, ApJ, 498, 579

\bibitem[ham]{mk87}
Hamilton, D., \& Keel, W.C. 1987, ApJ, 321, 211

\bibitem[iwa]{iwa99}
Iwasawa, K. 1999, MNRAS, 302, 961 

\bibitem[iwaco]{iwaco98}
Iwasawa, K., \& Comastri, A. 1998, MNRAS, 297, 1219 

\bibitem[iwafm]{ifm97}
Iwasawa, K., Fabian, A.C., \& Matt, G. 1997, MNRAS, 289, 443 

\bibitem[kk]{kk87}
Krolik, J.H., \& Kallman, T.R. 1987, ApJ, 320, L5

\bibitem[kmz]{kmz}
Krolik, J.H., Madau, P., \& \.Zycki, P.T. 1994, ApJ, 420, L57

\bibitem[lewmb]{lewmb}
Lawrence, A., Elvis, M., Wilkes, B.J., McHardy, I., \& Brandt,
N. 1997, MNRAS, 285, 879

\bibitem[mbf]{mbf}
Matt, G., Brandt, W.N., \& Fabian, A.C. 1996, MNRAS, 280, 823

\bibitem[mm90]{mm90}
Melnick, J., \& Mirabel, I.F. 1990, A\&A, 231, L19

\bibitem[riek88]{riek88}
Rieke, G.H. 1988, ApJ, 331, L5

\bibitem[rlrr]{rlrr}
Rigopoulou, D., Lawrence, A., \& Rowan-Robinson, M. 1996, MNRAS, 278,
1049 

\bibitem[sandm]{sm96}
Sanders, D.B., \& Mirabel, I.F. 1996, ARAA, 34, 749

\bibitem[sandet]{sandet88}
Sanders, D.B., Soifer, B.T., Elias, J.H., Madore, B.F., Matthews, K.,
Neugebauer, G., \& Scoville, N.Z. 1988, ApJ, 325, 74 

\bibitem[smithy]{ssaa95}
Smith, P.S., Schmidt, G.D., Allen, R.G., \& Angel, J.R.P. 1995, ApJ,
444, 146 

\bibitem[soif]{soif84}
Soifer, B.T. \etal\ 1984, ApJ, 278, L71

\bibitem[sturmet]{sturmet96}
Sturm, E., \etal\ 1996, A\&A, 315, L133

\bibitem[tur]{tur99}
Turner, T.J. 1999, ApJ, 511, 142

\bibitem[vignet]{vignet99}
Vignati, P., \etal\ 1999, A\&A, 349, L57

\bibitem[war]{ward88}
Ward, M.J., Done, C., Fabian, A.C., Tennant, A.F., \& Shafer,
R. A. 1988, ApJ, 324, 767

\end{thebibliography}
\end{document}